\documentclass[preprint, aps, prd, nofootinbib, floatfix]{revtex4-1}
\usepackage{graphicx} 
\usepackage{amsmath, amssymb, bm}
\usepackage{hyperref}
\usepackage{siunitx}
\usepackage{booktabs}
\usepackage{array}

\newcommand{\FIRSTAFF}{\affiliation{Department of Physics, University at Buffalo, Buffalo, NY 14260, USA}}

\usepackage{ulem}
\usepackage{tikz}

\begin{document}

\title{Phase Transitions and Gravitational Waves}
\author{Diego Rios}
\email[Electronic address: ]{diegorio@buffalo.edu}
\FIRSTAFF
\author{William H. Kinney}
\email[Electronic address: ]{whkinney@buffalo.edu}
\FIRSTAFF
\date{July 2026}

\begin{abstract}
We present a Fisher matrix forecast for the spectral parameter reconstruction of a stochastic gravitational wave background generated by a first-order phase transition in the early universe. We use the LISA and DECIGO instruments for reference and model the source spectrum with a double-broken-power-law template, parameterized by the peak amplitude $\Omega_p$, peak frequency $f_p$, break ratio $r_b$, and intermediate slope $b$. For each detector, we construct a fixed fiducial benchmark with the peak placed in the corresponding sensitivity band, using $f_p=3.0\times10^{-3} \ {\rm Hz}$ for LISA and $f_p=1.0\ {\rm Hz}$ for DECIGO. The LISA baseline includes an unresolved galactic compact-binary foreground, while for DECIGO we compare foreground cases using a full compact binary, a projected post-subtraction compact-binary residual, and an instrument noise only foreground. We find that both detectors constrain the peak parameters $\ln\Omega_p$ and $\ln f_p$ more robustly than the detailed shape parameters $r_b$ and $b$. For the chosen DECIGO benchmark, the projected residual foreground produces negligible parameter degradation relative to the instrument noise only case, while the full foreground lowers the signal-to-noise ratio from $\rho=10$ to $\rho\simeq6.06$ and increases the marginalized uncertainty on $b$ by about $51\%$. These results emphasize that stochastic-background forecasts should distinguish detectability from spectral identifiability.
\end{abstract}

\maketitle

\section{Introduction}
A first-order phase transition in the early universe can generate a stochastic gravitational-wave background through the dynamics of expanding bubbles and the subsequent motion of the surrounding plasma. The resulting spectrum is not only a potential discovery channel for physics beyond the Standard Model, but also a source of information about the characteristic time scale, energy scale, and dynamics of the transition. A cosmological first-order phase transition proceeds through the nucleation and expansion of bubbles of the stable phase within a metastable background. Gravitational waves can be sourced by scalar field dynamics associated with bubble collisions, acoustic motion in the surrounding plasma, and magnetohydrodynamic turbulence. In physical phase transition models, the amplitude, characteristic frequencies, and shape of the resulting spectrum depend on quantities such as the transition strength $\alpha$, inverse duration $\beta/H_*$, transition temperature $T_*$, and wall velocity $v_w$ \cite{Caprini:2019egz}. Future space-based gravitational wave detectors such as LISA and DECIGO therefore provide probes of early universe phase transitions. Forecasts formulated in terms of physical transition parameters can exhibit substantial parameter degeneracies because different combinations may produce similar spectral shapes. Realistic spectral models account for the dependence of the spectral shape on the underlying physical parameters and can therefore partially lift these degeneracies \cite{Gowling:2022pzb}. Modern LISA reconstruction pipelines have further studied the connection between phenomenological spectra, detector-level reconstruction, and the interpretation of the underlying phase-transition physics \cite{Caprini:2024hue}. For this reason, it is useful to first ask a more direct observational question. Which spectral features of the stochastic background can a detector actually measure?  Following the double-broken-power-law description used in previous LISA studies \cite{Gowling:2021gcy}, we parameterize the spectrum by its peak amplitude $\Omega_p$, peak frequency $f_p$, break ratio $r_b$, and intermediate slope $b$. Fisher forecasts for phase transition spectra with LISA, DECIGO, and BBO have been studied previously, but did not focus on the impact of a DECIGO specific compact-binary subtraction residual on spectral parameter recovery \cite{Hashino:2018wee}. We revisit this type of analysis using the double-broken-power-law spectral parameterization, focusing on spectral identifiability rather than the direct reconstruction of thermodynamic parameters. We incorporate recently developed compact-binary foreground-subtraction estimates for DECIGO based on binary populations inferred from recent LIGO--Virgo--KAGRA observations \cite{Yamamoto:2026air}. This makes it important to distinguish between instrumental noise, projected post-subtraction residuals, and the full unsubtracted compact-binary foreground when forecasting DECIGO sensitivity. In this work, we present a fixed-fiducial Fisher analysis of double-broken-power-law spectral recovery for LISA and DECIGO. For LISA, we include the unresolved Galactic compact-binary foreground in the baseline effective noise model. For DECIGO, we compare three foreground assumptions: instrument noise only, a projected post-subtraction compact-binary residual, and the full compact-binary foreground. This setup allows us to quantify which spectral parameters are measurable and how DECIGO compact-binary foreground assumptions affect both the signal-to-noise ratio and the marginalized Fisher uncertainties.

\section{Methodology}\label{Methods}
\subsection{Phase Transition source spectrum}
%this is dbpl source model from peer review
In this analysis, the stochastic gravitational-wave background sourced by a first-order phase transition is modeled for both LISA and DECIGO using the double-broken-power-law template from Ref. \cite{Gowling:2021gcy}
\begin{equation}
    \Omega_{\rm GW}(f) =
    \Omega_{p} \ M(s , r_b , b),
\end{equation}
where
\begin{equation}
    s = \frac{f}{f_p}.
\end{equation}
Here $\Omega_p$ is the peak amplitude of the spectrum and $f_p$ is the 
peak frequency. The parameter $r_b$ is
\begin{equation}
    r_b = \frac{f_b}{f_p},
\end{equation}
where $f_b$ is the lower break frequency. This template parametrizes a continuous spectrum with different approximate power-law regimes. The low-frequency and high-frequency asymptotic slopes are fixed to $9$ and  $-4$, respectively, so the parameter $b$ controls the intermediate spectral slope between the lower break and the peak region. $M(s , r_b, b)$ is a shape function defined as
\begin{equation}
    M(s , r_b , b) = s^9 \left(\frac{1 +r_b^4}{r_b^4 + s^4}\right)^{(9-b)/4} \ \left( \frac{b + 4}{b + 4 - m + ms^2} \right)^{(b + 4) / 2},
\end{equation}
with
\begin{equation}
    m = \frac{9 r_b^4 + b}{r_b^4 + 1}.
\end{equation}
The value of $m$ is chosen so that for $r_b < 1$, the template is normalized according to $M(1, r_b, b) = 1$, making the spectrum peak at $f = f_p$ with amplitude
\begin{equation}
    \Omega_{\rm GW}(f_p) = \Omega_p.
\end{equation}
The spectral parameter vector used in the Fisher analysis is therefore
\begin{equation}\label{Eq: spectral param vec}
    \theta_{\rm spec} = (\ln{\Omega_p} , \ln{f_p} , r_b , b ).
\end{equation}

\subsection{LISA Sensitivity and forecast signal-to-noise ratio}
 We take the effective LISA strain sensitivity from Ref. \cite{Robson:2018ifk} and augment it with an unresolved galactic binary foreground rather than reproduce the full A/E time delay interferometry likelihood from Ref. \cite{Gowling:2021gcy}. The noise model for the LISA instrument is thus
\begin{equation}\label{LISA Sen eq 13}
    S_n^{\text{LISA}}(f) = \frac{10}{3L^2}\left(P_{\rm OMS}+2(1 + \cos^2({f/f_*}))\frac{P_{\rm acc}}{(2\pi f)^4} \right)\left(1 + \frac{6}{10}\left(\frac{f}{f_*}\right)^2\right),
\end{equation}
where for $f_* = 19.09 \ \text{mHz}$ and $L =2.5 \ \text{Gm}$, the optical metrology noise $P_{\rm OMS}$ is
\begin{equation}
    P_{\rm OMS} = (1.5 \times10^{-11} \ \text{m})^2 \left(1 + \left(\frac{2 \ \text{mHz}}{f}\right)^{4}\right)\text{Hz}^{-1}.
\end{equation}
The acceleration noise $P_{\rm acc}$ is
\begin{equation}
    P_{\rm acc} = 
    (3 \times10^{-15} \text{m  s}^{-2})^2\left(1 + \left(\frac{0.4 \ \text{mHz}}{f}\right)^2\right)\left(1 + \left(\frac{f}{8 \ \text{mHz}}\right)^4\right)\text{Hz}^{-1}.
\end{equation}
The effective LISA strain-noise power spectral density used in the forecast is
\begin{equation}
P_{\rm LISA}(f) =
S_{\rm n}^{\rm LISA}(f) +
S_{\rm gb}(f),
\end{equation}
where $S_{\rm gb}(f)$ is the galactic binary foreground from Ref. \cite{Gowling:2021gcy}
\begin{equation}
S_{\rm gb}(f) = A_{\rm gb}f^{-7/3} \exp\left[ -f^{a_{\rm gb}}
+ b_{\rm gb} f \sin(c_{\rm gb} f) \right] \left[ 1+\tanh\left( d_{\rm gb}(f_k-f) \right)
\right].
\end{equation}
The values of the numerical coefficients are taken as follows: $A_{\rm gb} = 9 \times10^{-45}$, $a_{\rm gb} = 0.138$, $b_{\rm gb} = -221$, $c_{\rm gb} = 521$, $d_{\rm gb} = 1680$, and $f_{k} = 1.13 \times 10^{-3} \  \text{Hz}$. The double-broken-power-law energy density spectrum is converted to an effective strain spectrum using
\begin{equation}\label{Eqn: cov-to-strain}
S_h^{\rm LISA}(f)
=
\frac{3H_0^2}{10\pi^2}
\frac{\Omega_{\rm GW}(f)}{f^3},
\end{equation}
for
\begin{equation}
H_0
=
67.4 \,
{\rm km}\,
{\rm s}^{-1}\,
{\rm Mpc}^{-1}.
\end{equation}
The forecast signal-to-noise ratio is computed as
\begin{equation}
\rho_{\rm LISA}^2
=
2T_{\rm obs}N_{\rm ch}
\int_{f_{\rm min}}^{f_{\rm max}}
df \ 
\frac{
\gamma^2(f)
\left[
S_h^{\rm LISA}(f)
\right]^2
}{
P_{\rm LISA}^2(f)
}.
\end{equation}
In the LISA forecasts presented here, we take observation time, $T_{\rm obs} = 4 \text{yrs}$, overlap reduction function $\gamma(f)=1$, and use a single effective channel, $N_{\rm ch}=1$. The implementation therefore corresponds to a forecast-level effective noise Fisher analysis.

\subsection{DECIGO Sensitivity and Forecast Signal-to-Noise Ratio}
Similarly, for DECIGO the gravitational-wave energy density is converted to an effective strain spectrum using
\begin{equation}
S_h^{\rm DECIGO}(f)
=
\frac{3H_0^2}{10\pi^2}
\frac{\Omega_{\rm GW}(f)}{f^3}.
\end{equation}
We model the DECIGO instrument noise using the analytic fit from Ref. \cite{Yagi:2011wg}, which gives the strain-noise power spectral density used for each effective DECIGO channel as
\begin{equation}
\begin{aligned}
S_n^{\rm DECIGO}(f)
&=
7.05\times 10^{-48}
\left[
1+
\left(
\frac{f}{f_d}
\right)^2
\right]
\\
&\quad
+
4.8\times 10^{-51}
\left(
\frac{f}{1\,{\rm Hz}}
\right)^{-4}
\frac{1}{
1+
\left(
\frac{f}{f_d}
\right)^2
}
\\
&\quad
+
5.33\times 10^{-52}
\left(
\frac{f}{1\,{\rm Hz}}
\right)^{-4}
\,{\rm Hz}^{-1},
\end{aligned}
\end{equation}
with
\begin{equation}
f_d = 7.36 \ {\rm Hz}.
\end{equation}
We introduce a foreground contribution from binary neutron star mergers generated using Ref. \cite{Yamamoto:2026air}. Foreground contributions are implemented through the following equation for strain-noise power spectral density
\begin{equation}
P_{\rm DECIGO}(f)
=
S_n^{\rm DECIGO}(f)
+
S_h^{\rm fg}(f),
\end{equation}
where 
\begin{equation}\label{Eqn: cov-to-strain-DECIGO}
S_h^{\rm fg}(f)
=
\frac{3H_0^2}{10\pi^2}
\frac{\Omega_{\rm fg}(f)}{f^3}.
\end{equation}
We consider three foreground choices:
\begin{equation}
\Omega_{\rm fg}(f)
=
\begin{cases}
0,
&
\text{instrument-specific noise only},
\\
\Omega_{\rm Full}(f),
&
\text{total compact-binary foreground},
\\
\Omega_{\rm Err(projected)}(f),
&
\text{projected waveform-subtraction residual}.
\end{cases}
\end{equation}
Here ``projected waveform-subtraction residual'' 
refers to the foreground left after individually detectable compact-binary signals are modeled and their best-fit waveforms are subtracted from the strain data. The projected residual applies the Cutler--Harms projection scheme to suppress the leading parameter estimation error contribution to the subtraction residual. For DECIGO, we compute the signal-to-noise ratio as
\begin{equation}
\rho_{\rm DECIGO}^2
=
2 T_{\rm obs} N_{\rm ch}
\int_{f_{\rm min}}^{f_{\rm max}}
df\,
\frac{
\gamma^2(f)
\left[
S_h^{\rm DECIGO}(f)
\right]^2
}{
P_{\rm DECIGO}^2(f)
},
\end{equation}
for the following values
\begin{equation}
T_{\rm obs}=4\,{\rm yr},
\qquad
f_{\rm min}=0.01\,{\rm Hz},
\qquad
f_{\rm max}=100\,{\rm Hz},
\end{equation}
and assume
\begin{equation}
\gamma(f)=1.
\end{equation}
DECIGO is usually associated with a smaller frequency range, the decihertz band, but we use this broader range because the analytic noise fit from Ref. \cite{Yagi:2011wg} is evaluated over this broader interval. Additionally, the double-broken-power-law template has extended power-law tails away from its peak. We use $N_{\rm ch}=8$, matching the convention used to generate the foreground source file. This factor approximates the quadrature sum over statistically independent DECIGO cross-correlation channels.

\subsection{Fiducial spectral model and amplitude normalization}
For the benchmark spectral shape we take
\begin{equation}
r_b = 0.3,
\qquad
b = 1.0.
\end{equation}
The fiducial peak frequency is chosen separately for each detector so that the signal lies in the relevant sensitivity band. For LISA we use
\begin{equation}
f_p = 3.0\times10^{-3} \ {\rm Hz},
\end{equation}
and for DECIGO we use
\begin{equation}
f_p = 1.0 \ {\rm Hz}.
\end{equation}
The peak amplitude is then fixed initially for each detector-centered benchmark. For each instrument we use its baseline noise model to determine which amplitude values will give a relatively high signal-to-noise ratio $\rho = 10$. For LISA, this value is
\begin{equation}
\Omega_p = 8.50\times10^{-13},
\end{equation}
if the noise model also includes the contribution from the unresolved Galactic binary foreground. For DECIGO the value is
\begin{equation}
\Omega_p = 3.36\times10^{-15},
\end{equation}
if we consider no foreground contributions other than the noise of the instrument in the determination of this amplitude. After this normalization, the fiducial double-broken-power-law signal is held fixed. In particular, for the DECIGO foreground comparisons, the same values of $\Omega_p$, $f_p$, $r_b$, and $b$ are used in the instrument only, projected residual, and full compact-binary foreground cases. Therefore, changes in the recovered signal-to-noise ratio and marginalized uncertainties are caused only by changes in the assumed detector foreground model. We do not impose any priors on the Fisher matrices presented here.

\subsection{Fisher matrix}
For both LISA and DECIGO, the Fisher matrix is computed using the same effective-noise approximation. Since the single strain power spectrum is
\begin{equation}
S_h^{\rm GW}(f)
=
\frac{3H_0^2}{10\pi^2}
\frac{\Omega_{\rm GW}(f)}{f^3},
\end{equation}
the Fisher matrix in this forecast follows from a Gaussian approximation to the likelihood for a cross-correlation estimator of this single strain power spectrum. We set the estimator equal to the fiducial signal
\begin{equation}
\hat{S}_h(f) = S_h^{\rm GW}(f;\boldsymbol{\theta}_0),
\end{equation}
so that the Fisher matrix describes the local curvature of the likelihood around the fiducial spectral model. For these matrices, the instrumental noise and foreground contributions are held fixed while the spectral parameters in Eq. \ref{Eq: spectral param vec} are varied. For LISA, the Fisher matrix is
\begin{equation}
F_{ij}^{\rm LISA}
=
2T_{\rm obs}N_{\rm ch}
\int_{f_{\rm min}}^{f_{\rm max}} df\,
\frac{\gamma^2(f)}
{
P_{\rm LISA}^2(f)
}
\left.
\frac{\partial S_h^{\rm GW}(f)}{\partial\theta_i}
\frac{\partial S_h^{\rm GW}(f)}{\partial\theta_j}
\right|_{\boldsymbol{\theta}=\boldsymbol{\theta}_0},
\end{equation}
and for DECIGO the Fisher matrix is
\begin{equation}
F_{ij}^{\rm DECIGO}
=
2T_{\rm obs}N_{\rm ch}
\int_{f_{\rm min}}^{f_{\rm max}} df\,
\frac{\gamma^2(f)}
{
P_{\rm DECIGO}^2(f)
}
\left.
\frac{\partial S_h^{\rm GW}(f)}{\partial\theta_i}
\frac{\partial S_h^{\rm GW}(f)}{\partial\theta_j}
\right|_{\boldsymbol{\theta}=\boldsymbol{\theta}_0}.
\end{equation}
The parameter covariance matrix is approximated by the inverse Fisher matrix,
\begin{equation}
\Sigma_{ij}
=
\left(F^{-1}\right)_{ij} \ ,
\end{equation}
and the marginalized $1\sigma$ uncertainty on parameter $\theta_i$ is
\begin{equation}
\sigma(\theta_i)
=
\sqrt{\Sigma_{ii}}.
\end{equation}
This Fisher implementation differs from the one used in Ref. \cite{Gowling:2021gcy}. In that work, the LISA likelihood is written in terms of the A/E time delay interferometry channels. For this analysis we use a common effective noise approximation for both LISA and DECIGO in which the detector noise and foreground terms are treated as fixed. This corresponds to the standard weak-signal, or noise-dominated, approximation used in stochastic background cross-correlation forecasts. In this limit, the signal contribution to the variance of the estimator is neglected relative to the detector noise and the foreground. As such, this approximation may not be valid in cases involving arbitrarily loud backgrounds. Ref. \cite{Liang:2024ulf} shows that this approximation may overestimate the signal-to-noise ratio in the strong signal regime where the stochastic gravitational wave background power becomes comparable to the detector noise. Our forecasts therefore are restricted to moderate signal-to-noise ratio, but more importantly, to cases in which the signal power remains well below the effective noise power. To check the consistency of the weak-signal approximation for the chosen benchmarks, we compute the ratio
\begin{equation}
    R(f)
    =
    \frac{S_h^{\rm GW}(f)}{P_{\rm eff}(f)} .
\end{equation}
Here $P_{\rm eff}$ denotes $P_{\rm LISA}$ or $P_{\rm DECIGO}$ for the corresponding detector and foreground model. For the LISA fiducial signal, the maximum value is $R_{\rm max}=1.47\times10^{-2}$. For DECIGO, the maximum values are $R_{\rm max}=4.02\times10^{-4}$ if the instrument-noise-only foreground and projected residual-foreground are considered individually. $R_{\rm max}=2.21\times10^{-4}$ for the full DECIGO foreground case. Since $S_h^{\rm GW}/P_{\rm eff}\ll 1$ throughout the Fisher-weighted frequency band for both LISA and DECIGO, the signal contribution to the variance is negligible for our benchmarks.

\section{Results}\label{results}
\begin{figure}
    \centering
    \includegraphics[width=0.8\linewidth]{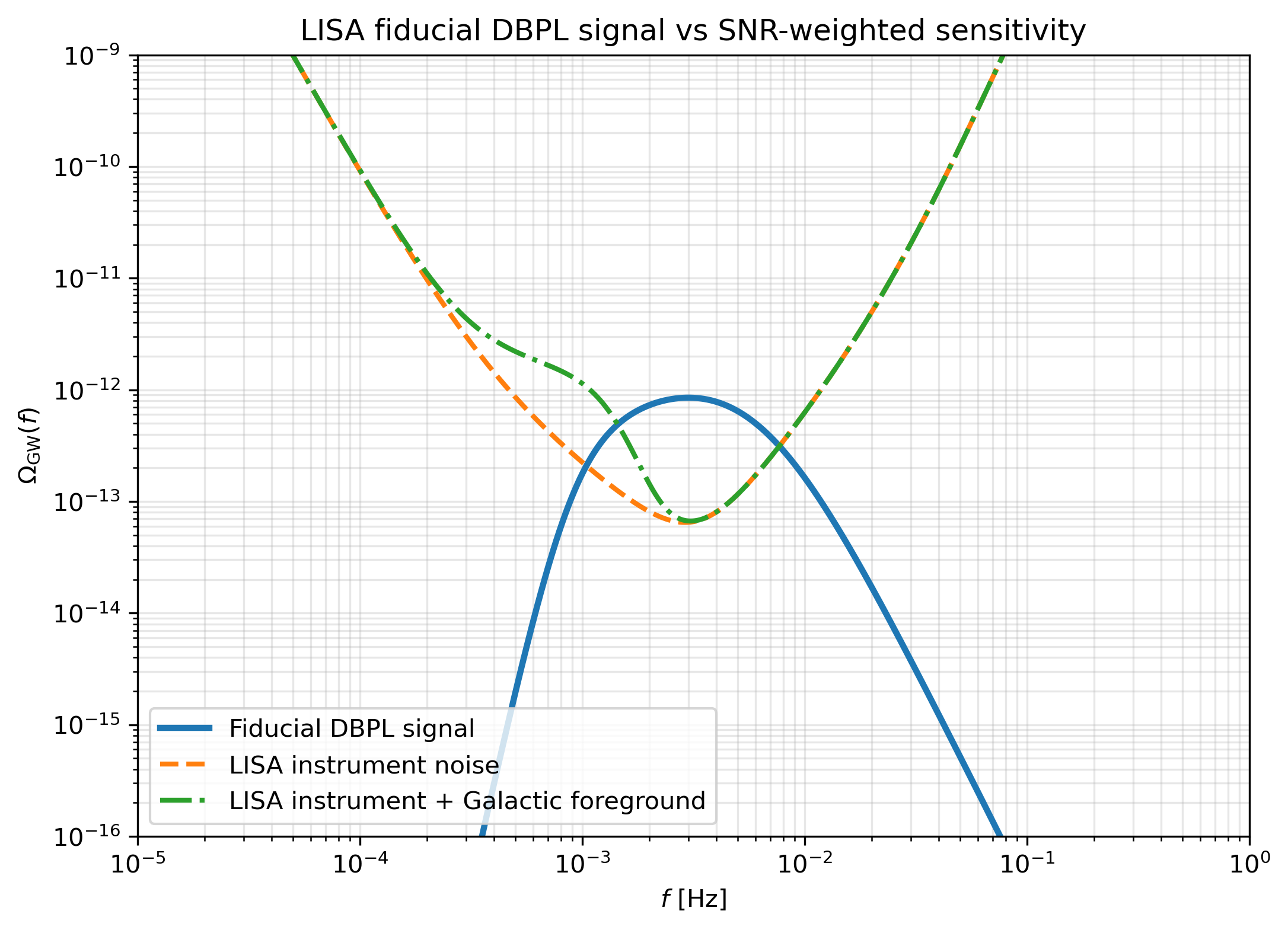}
    \caption{Fiducial LISA double-broken-power-law (DBPL) signal compared with the effective sensitivity curves used in the Fisher forecast. The sensitivity curves are shown in energy-density units for the instrument-only case and for the baseline model including the unresolved Galactic compact-binary foreground.}
    \label{fig: LISA amp_vs_signal}
\end{figure}

\begin{figure}
    \centering
    \includegraphics[width=0.8\linewidth]{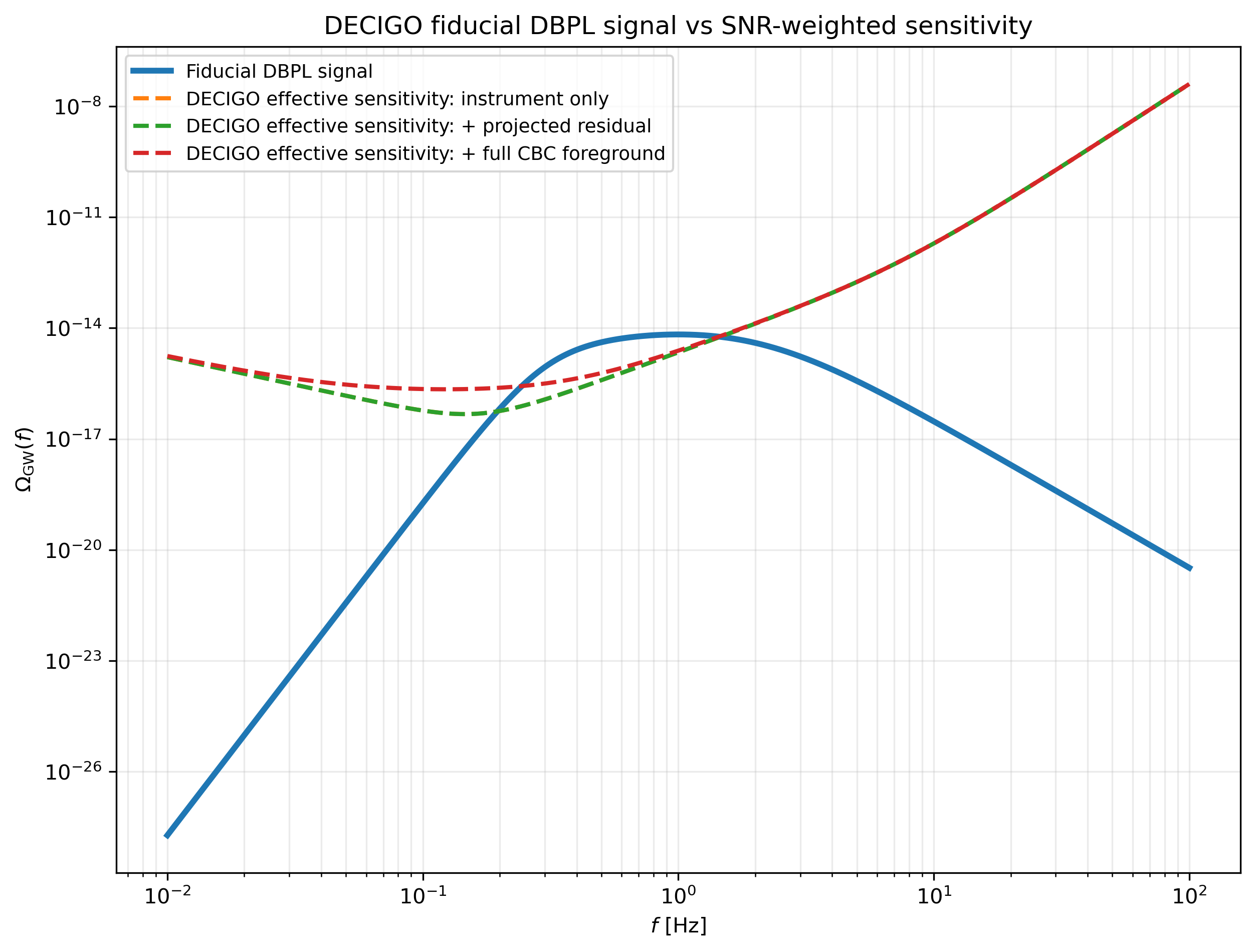}
    \caption{Fiducial DECIGO double-broken-power-law (DBPL) signal compared with the effective sensitivity curves used in the Fisher forecast. The sensitivity curves are shown in energy-density units and include the observation time, channel count, and per-log-frequency weighting entering the cross-correlation signal-to-noise ratio. The projected Yamamoto inspired residual foreground is nearly indistinguishable from the instrument-only curve, while the full compact-binary foreground visibly raises the effective sensitivity curve.}
    \label{fig: DECIGO amp_vs_signal}
\end{figure}

\begin{figure}
    \centering
    \includegraphics[width=0.85\linewidth]{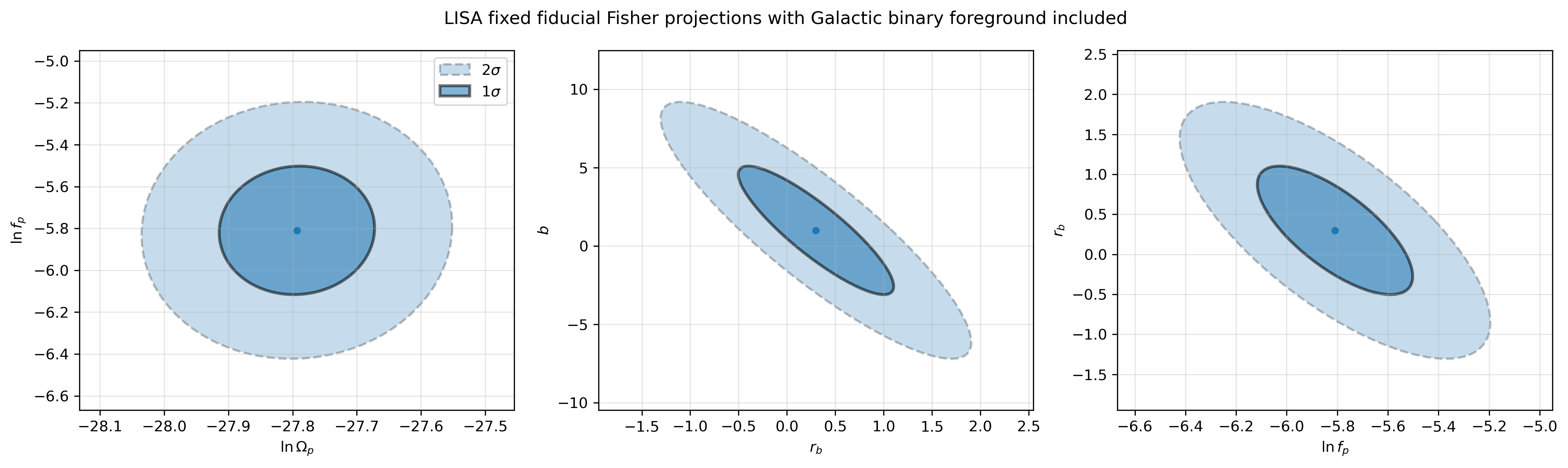}
    \caption{LISA Fisher projections for the fixed-fiducial double-broken-power-law benchmark with the unresolved Galactic binary foreground included in the baseline noise model.
    The fiducial signal has $f_p=3.0\times10^{-3}\,{\rm Hz}$, $r_b=0.3$, $b=1.0$, and $\Omega_p=8.50\times10^{-13}$, giving a signal-to-noise ratio of $\rho=10$. The panels show the marginalized uncertainties in the $(\ln\Omega_p,\ln f_p)$, $(r_b,b)$, and $(\ln f_p,r_b)$ planes. The inner and outer ellipses denote the $1\sigma$ and $2\sigma$ Fisher uncertainties, respectively. Table \ref{tab:lisa_fixed_fiducial_summary} summarizes the numerical results.}
    \label{fig: LISA ellipses}
\end{figure}

\begin{figure}
    \centering
    \includegraphics[width=0.85\linewidth]{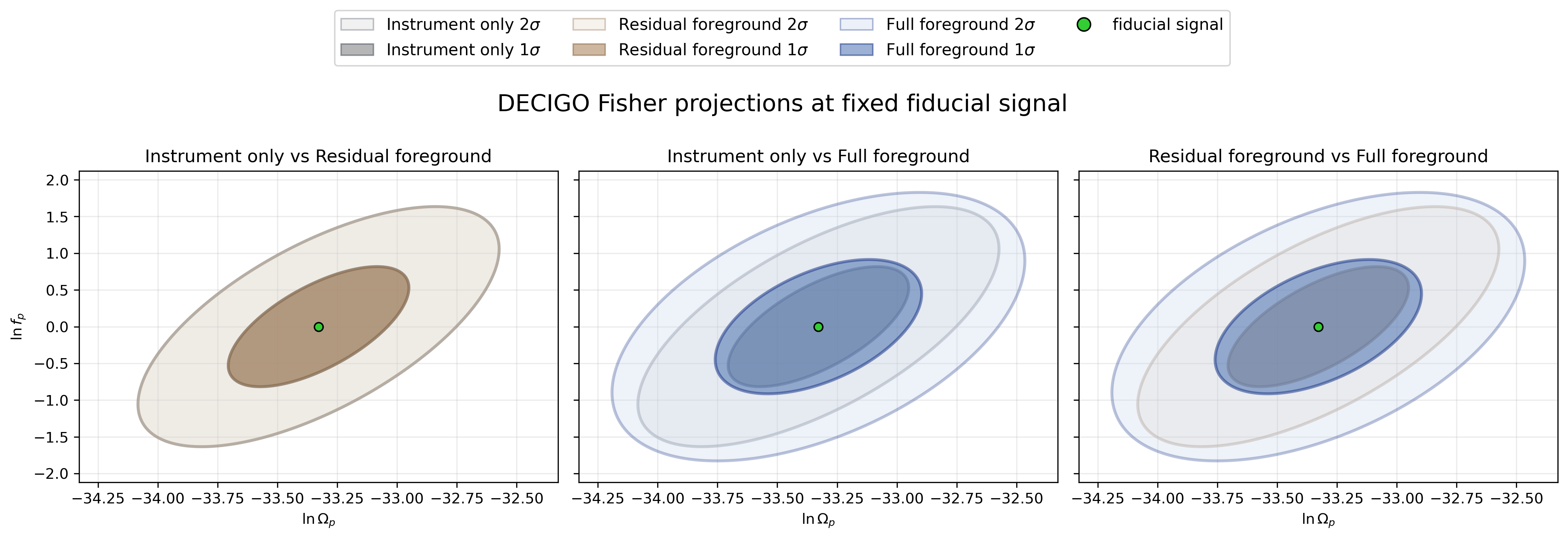}
    \caption{
    DECIGO Fisher projections in the $(\ln\Omega_p,\ln f_p)$ plane for a fixed
    fiducial double-broken-power-law signal. The same fiducial spectral parameters are used in all
    foreground cases and only the assumed DECIGO foreground contribution is changed. The three panels show comparisons between the instrument noise only case, the projected post-subtraction compact-binary residual, and the unsubtracted full compact-binary foreground. The inner and outer ellipses denote the $1\sigma$ and $2\sigma$ Fisher uncertainties, respectively. The projected residual foreground produces nearly indistinguishable constraints to the instrument only case, while the full foreground broadens the allowed region. The corresponding signal-to-noise ratio reduction and marginalized uncertainty increases are listed in Table \ref{tab:decigo_fixed_fiducial_foreground_degradation}.}
    \label{fig: DECIGO ellipses}
\end{figure}

For both LISA and DECIGO we provide signal amplitude vs frequency plots shown by Fig. \ref{fig: LISA amp_vs_signal} and Fig. \ref{fig: DECIGO amp_vs_signal} respectively. Fig. \ref{fig: LISA ellipses} and Fig. \ref{fig: DECIGO ellipses} show the marginalized Fisher uncertainties for the spectral parameters. Without any imposed priors, LISA weakly constrains the $r_b$ parameter, whose uncertainty extends into non-physical space when $r_b < 0$. The intermediate slope parameter $b$ is also poorly constrained, with a marginalized uncertainty much larger than the fiducial value $b=1$. By contrast, the parameters $\ln \Omega_{p}$ and $\ln{f_p}$ are best constrained, consistent with the claim reported in Ref. \cite{Gowling:2021gcy}. For this reason, the DECIGO parameter uncertainties for various foreground considerations in Fig. \ref{fig: DECIGO ellipses} are displayed only for the $\ln \Omega_{p}$ and $\ln{f_p}$ parameters, while the full set of marginalized uncertainties is reported in Table \ref{tab:decigo_fixed_fiducial_foreground_degradation}. For our fiducial benchmark, the DECIGO residual foreground and the instrument noise only foreground produce nearly identical parameter uncertainties, with differences of less than $1\%$. This suggests that the standard residual projected foreground from Ref. \cite{Yamamoto:2026air} does not significantly degrade parameter recovery. The full compact-binary foreground lowers the signal-to-noise ratio from the instrument-only reference value, $\rho = 10$, to $\rho \simeq 6.06$ and increases all the reported marginalized uncertainties. The largest effect is on $\sigma_b$, which increases by about $51\%$. This shows that the full unresolved/unsubtracted foreground affects not only detectability but also recovery of the spectral shape.
\begin{table}[htbp]
\centering
\caption{
LISA fixed fiducial Fisher summary for the double-broken-power-law benchmark. For this signal $f_p=3.0\times10^{-3}\,{\rm Hz}$, $r_b=0.3$, $b=1.0$, and $\Omega_p=8.50\times10^{-13}$. The baseline LISA noise model includes instrumental noise and the unresolved Galactic compact-binary foreground. The reported uncertainties are marginalized $1\sigma$ Fisher errors.}
\label{tab:lisa_fixed_fiducial_summary}
\begin{ruledtabular}
\scriptsize
\begin{tabular}{lccccc}
Foreground model
& $\rho$
& $\sigma_{\ln\Omega_p}$
& $\sigma_{\ln f_p}$
& $\sigma_{r_b}$
& $\sigma_b$ \\
\hline
Galactic binary baseline + instrument noise
& 10.000
& 0.121
& 0.307
& 0.803
& 4.10 \\
\end{tabular}
\end{ruledtabular}
\end{table}

\begin{table}[t]
\centering
\caption{
DECIGO fixed fiducial foreground-degradation summary for the double-broken-power-law benchmark.
The fiducial signal parameters are held fixed across all foreground assumptions.
The column $\rho$ gives the resulting signal-to-noise ratio for each foreground model, while $\Delta\rho/\rho_{\rm inst}$ gives the percent decrease in signal-to-noise ratio relative to the instrument only case. For a given spectral parameter $\theta_i$, the column $\sigma_{\theta_i}$ gives the marginalized $1\sigma$ Fisher uncertainty, and $\Delta\sigma_{\theta_i}$ gives the corresponding percent increase relative to the instrument only uncertainty.
}
\label{tab:decigo_fixed_fiducial_foreground_degradation}
\begin{ruledtabular}
\scriptsize
\begin{tabular}{lcccccccccc}
Foreground model
& $\rho$
& $\Delta\rho/\rho_{\rm inst}$
& $\sigma_{\ln\Omega_p}$
& $\Delta\sigma_{\ln\Omega_p}$ 
& $\sigma_{\ln f_p}$
& $\Delta\sigma_{\ln f_p}$ 
& $\sigma_{r_b}$
& $\Delta\sigma_{r_b}$ 
& $\sigma_b$
& $\Delta\sigma_b$  \\
\hline
Instrument noise only
& 10.000
& 0
& 0.377
& 0
& 0.817
& 0
& 0.220
& 0
& 2.93
& 0 \\

Residual foreground
& 10.000
& 0.00183
& 0.377
& 0.000302
& 0.817
& 0.000396
& 0.220
& 0.000308
& 2.93
& 0.00159 \\

Full foreground
& 6.059
& 39.4
& 0.431
& 14.3
& 0.914
& 11.9
& 0.255
& 16.0
& 4.42
& 51.0 \\
\end{tabular}
\end{ruledtabular}
\end{table}

\section{Conclusion}\label{conclusion}
In this work we presented a fixed-fiducial Fisher forecast for the spectral reconstruction of a stochastic gravitational wave background from a first-order phase transition. Rather than directly forecasting the thermodynamic parameters, we modeled the signal with a double-broken-power-law template and asked which spectral features are directly identifiable by LISA and DECIGO. Our results show that a stochastic background may be detectable while some of the parameters controlling its detailed shape remain weakly constrained. They also show that DECIGO foreground assumptions matter because a projected post-subtraction residual may be negligible for a particular benchmark, while a full unsubtracted compact-binary foreground can substantially degrade both the signal-to-noise ratio and the shape recovery. Future work should extend this approach in two possible directions. First, the analysis should be repeated over a wider range of peak frequencies and for different benchmark cases, including the microhertz gap emphasized by proposed concepts such as GUEST \cite{Blas:2026xol}. Second, the double-broken-power-law spectral covariance should be mapped back onto physical phase-transition parameters. Such extensions could clarify how much of the underlying early-universe physics can be recovered in concrete particle physics models. 

\section*{Acknowledgements}
This work is supported by the National Science Foundation under grant NSF-PHY-2310363.

\bibliography{references}

\end{document}